\documentclass[prl,twocolumn,showpacs,superscriptaddress]{revtex4}

\usepackage{bm}
\usepackage{graphicx}
\usepackage{times}
\usepackage[T1]{fontenc}

\newcommand{\be}{\begin{equation}}
\newcommand{\ee}{\end{equation}}
\newcommand{\ba}{\begin{eqnarray}}
\newcommand{\ea}{\end{eqnarray}}
\newcommand{\bsig}{{\mbox{\boldmath$\sigma$}}}
\newcommand{\sumk}{\int \! \frac{d^3k}{(2\pi)^3}\,}

\newcommand{\ka}{\mathbf{k}}
\newcommand{\q}{\mathbf{q}}

\newcommand{\pds}{(pd\sigma)}
\newcommand{\pdp}{(pd\pi)}

\begin{document}

\title{Influence of non-local exchange on RKKY interactions\\ in III-V
diluted magnetic semiconductors}

\author{C. Timm}
\email{timm@physik.fu-berlin.de}
\affiliation{Institut f\"ur Theoretische Physik, Freie Universit\"at Berlin,
Arnimallee 14, D-14195 Berlin, Germany}
\author{A. H. MacDonald}
\email{macd@physics.utexas.edu} 
\affiliation{Physics Department, University of Texas at Austin,
Austin TX 78712-0264}

\date{May 20, 2004}

\begin{abstract}
The RKKY interaction between substitutional Mn local moments in GaAs is
both spin-direction-dependent and spatially anisotropic. In this Letter we
address the  strength of these anisotropies using a semi-phenomenological
tight-binding  model which treats the hybridization between Mn
d-orbitals and As p-orbitals  perturbatively and accounts realistically for
the non-local exchange interaction between their spins.  We show that
exchange non-locality, valence-band spin-orbit coupling, and band-structure
anisotropy all play a role in determining the strength of both effects. 
We use these results to estimate the degree of ground-state magnetization
suppression due to frustrating interactions between randomly located Mn
ions. 
\end{abstract}


\maketitle

The current interest in diluted magnetic semiconductors (DMS) is fueled by
possible applications in spintronics and by basic-science issues generated 
by the interplay between disorder, spin-orbit coupling, and  magnetic
order. We concentrate on the prototypical III-V DMS ${\rm Ga}_{1-x}{\rm
Mn}_{x}{\rm As}$, which, once interstitial Mn ions have been eliminated,
exhibits robust homogeneous ferromagnetism \cite{DMSFerro} with critical
temperatures $T_c$ above $160\:{\rm K}$ for $x \gtrsim 0.05$. It is
generally agreed that the substitutional Mn ions are in ${\rm Mn}^{2+}$
valence states that have $S=5/2$, $L=0$ local moments, and that exchange
interactions with As neighbors allow the Mn moments to interact via
valence-band holes \cite{holemediated}. The effective exchange interaction
between Mn moments is spatially anisotropic and, because of spin-orbit
interactions, also anisotropic in spin space. This Letter is motivated
primarily by theoretical interest \cite{ZaJ02,BrG03} in the role of
anisotropies in determining the character of the magnetic ground state but
has implications for other aspects of (III,Mn)V DMS ferromagnetism.

The theory of (III,Mn)V ferromagnetism has been developed in several
directions. A simple phe\-no\-me\-no\-lo\-gi\-cal approach
\cite{DHM97,JAL99,DOM00,KSJ03,AJB01,DOM01,SKM01} approximates the
valence-band holes by a host-semiconductor Kohn-Luttinger envelope-function
Hamiltonian and couples them to randomly located Mn spins by a
\emph{local}, \emph{isotropic} exchange interaction $J_{\mathrm{pd}}$. This
leads to a semi-quantitative description of many transport and magnetic
properties, particularly in the high-carrier-density, high-$T_c$ systems
that are free of compensating Mn interstitials. However, it has led to
conflicting conclusions  on the importance of exchange anisotropy. The RKKY
interaction obtained by Zar\'and and Jank\'o \cite{ZaJ02} is highly
anisotropic in \emph{spin} space, {\it i.e.}, it depends strongly on the
orientation of  two spins relative to their connecting vector, but it is
spatially isotropic because it starts from a {\em local} hole-impurity
exchange interaction and uses a {\em spherical} approximation for the
bands. Using a more realistic 6-band envelope-function Hamiltonian, Brey
and G\'omez-Santos \cite{BrG03} find that both spin and real space
anisotropies are weak. Their conclusion, however, depends in part on their
momentum-space cut-off \cite{SKM01} for the exchange interaction
$J_{\mathrm{pd}}$, {\it i.e.}, on atomic-length-scale physics not described
realistically in the envelope-function approach. First-principles
calculations \cite{LSDA} do not have these limitations, but are hampered by
their extreme sensitivity to the placement of unoccupied and occupied
d-orbital energies relative to the valence and conduction bands.  In this
Letter we address exchange anisotropy using a realistic tight-binding model
that combines virtues of these two different approaches and estimate the
bulk magnetization suppression due to frustrating interactions  between
impurity moments. Based on our results we also suggest a possible route
toward higher transition temperatures in (III,Mn)V ferromagnets.

Our theory is based on a Slater-Koster \cite{SK}  tight-binding model,
and on a perturbative treatment of pd hybridization, in which the band
electrons are integrated out to yield a spin-only  model
\cite{DHM97,ZaJ02,BrG03,PHS03,BKB03,Trev}. A similar model
has recently been used to obtain the local density of states around Mn
impurities \cite{TaF03}. In Slater-Koster theory, the electronic structure
is specified  by orbital-dependent onsite energies and hopping amplitudes
that are treated as fitting parameters. Spin-orbit coupling is
included \cite{Chadi} to obtain realistic bands and a realistic description
of (III,Mn)V ferromagnetism \cite{KSJ03}.  

Our Hamiltonian reads
$H = H_c + H_d + H_{\mathrm{hyb}}$, where
\be
H_c = \sum_{\ka} \sum_{\alpha\alpha'\sigma\sigma'}
  \epsilon_{\alpha\sigma;\alpha'\sigma'}(\ka)\, c_{\ka\alpha\sigma}^\dagger
  c_{\ka\alpha'\sigma'}
\ee
describes perfect GaAs \cite{SK,Chadi}.
Here, $c_{\ka\alpha\sigma}^\dagger$ creates an electron with wave vector
$\ka$ in orbital $\alpha$ with spin $\sigma$.
The most important effect of Mn impurities is to introduce
partially filled d-orbitals. The resulting strong electron-electron
interactions are parametrized by the local
Hubbard repulsion $U$ and the Hund's-first-rule coupling $J_H$
\cite{Parm,Kac01}:
$H_d = (\epsilon_d + J_H
- {U}/{2}) \hat N
+ {1}/{2}\,(U-{J_H}/{2}) \hat N^2
- J_H\, \mathbf{S}\cdot\mathbf{S}$,
with $\hat N\equiv \sum_{n\sigma} d_{n\sigma}^\dagger
d_{n\sigma}$ and $\mathbf{S} \equiv \sum_{n\sigma\sigma'}
d_{n\sigma}^\dagger\, ({\bsig_{\sigma\sigma'}}/{2})\, d_{n\sigma'}$, where
$d_{n\sigma}^\dagger$ creates an electron in d-orbital $n$ with spin
$\sigma$. We assume $U\approx 3.5\:\mathrm{eV}$ \cite{Oka} and
$J_H\approx 0.55\:\mathrm{eV}$ \cite{Craco}. 
$H_{\mathrm{hyb}}$ describes the hybridization between the
d-orbitals and sp-bands,
\be
H_{\mathrm{hyb}} = \frac{1}{\sqrt{\cal N}} \sum_{\ka}
  \sum_{\alpha\sigma n}
  t_{\ka\alpha n}\, c_{\ka\alpha\sigma}^\dagger d_{n\sigma}
  + \mathrm{h.c.}
  \equiv H_{\mathrm{hyb}}^- +  H_{\mathrm{hyb}}^+ ,
\label{Hhyb2}
\ee
where ${\cal N}$ is the number of unit cells in the system. The
coefficients are expressed in terms of real-space hopping matrix elements,
$t_{\ka\alpha n} = \sum_i e^{-i\ka\cdot\mathbf{u}_i} t_{i\alpha n}$, where
the sum runs over nearest-neighbor As sites of the impurity. The symmetries
of $t_{\ka\alpha n}$ are obtained from Slater-Koster theory \cite{SK},
which expresses the matrix elements in terms of two-center integrals. We
use $(pd\sigma)=1.0\:\mathrm{eV}$ and $(pd\pi)=-0.46\:\mathrm{eV}$ as
inferred  from photoemission \cite{Oka} and $(sd\sigma)=1.5\:\mathrm{eV}$
obtained as a rough spin average of \textit{ab-initio} calculations for
zinc-blende MnAs \cite{SaH00}.

In the large-$U$ limit we can use canonical perturbation theory (CPT)
\cite{SWC} to integrate out d-shell charge fluctuations, leaving only the
impurity spin degrees of freedom. We first consider a single Mn impurity.
We introduce the canonically transformed Hamiltonian $\tilde H \equiv
e^{-i\epsilon T}\, (H_c + H_d + \epsilon H_{\mathrm{hyb}})\, e^{i\epsilon
T}$, where $T$ is hermitian, and expand in $\epsilon$. The operator $T$ is
chosen so that the linear term vanishes.  To  obtain manageable expressions
we neglect the energetic spread of virtual band-electron states  compared
to the energy difference $\sim U$ between different Mn valence states. To
be consistent we ignore contributions from bands other than the 
heavy-hole, light-hole, and split-off bands. Truncating the expansion at
second order and projecting onto the $N=5$, $S=5/2$ ground-state subspace,
we obtain
\be
\tilde H \cong H_c +
  \frac{H_{\mathrm{hyb}}^+ H_{\mathrm{hyb}}^-}
  {E_{5,5/2}-E_{4,2}}
  + \frac{H_{\mathrm{hyb}}^- H_{\mathrm{hyb}}^+}
  {E_{5,5/2}-E_{6,2}} .
\label{2.PHP6}
\ee
We have used that $H_{\mathrm{hyb}}^\pm$ applied to a state in the
$(N,S)=(5,5/2)$ sector results in a state with sharp quantum numbers
$(N,S)=(6,2)$ and $(4,2)$, respectively. $E_{NS}$ is the corresponding
isolated-ion energy.
Inserting Eq.~(\ref{Hhyb2})
and noting that $\sum_{\sigma\sigma'} d_{n\sigma}^\dagger\,
(\bsig_{\sigma\sigma'}/2)\, d_{n\sigma'}
=\mathbf{S}/5$ in the $(5,5/2)$ sector, we obtain
a Hamiltonian that includes a microscopic hole-impurity
exchange interaction, 
\ba
\lefteqn{
\tilde H \:=\: H_c \:+\: \mbox{(charge scattering)} }
  \nonumber \\
& & \!\!\!{}- \frac{1}{\Delta}\,
  \frac{1}{\cal N} \sum_{\ka,\ka'} \sum_{\alpha\alpha'n}
  t^\ast_{\ka\alpha n} t_{\ka'\alpha'n} \sum_{\sigma\sigma'}
  c^\dagger_{\ka'\alpha'\sigma'}\frac{\bsig_{\sigma'\sigma}}{2}
  c_{\ka\alpha\sigma} \cdot \mathbf{S} \hspace{1.5em}
\label{2.Htrans2}
\ea
with
\be
\frac{1}{\Delta} \equiv \frac{2}{5}\,
  \bigg(\frac{1}{\epsilon_d-4J_H+4U}
  + \frac{1}{-\epsilon_d-J_H-5U}\bigg) .
\ee
The two energy denominators in $1/\Delta$
are respectively the isolated-ion $d^5\!\to\!d^4$ and $d^5\!\to\!d^6$
transition energies measured from the chemical potential. If either of
the denominators becomes small, the interval of energy over which
our approximations are justified is correspondingly reduced.
Note first that the exchange interaction is quite generally invariant
under spin rotation. The wavevector dependence of the exchange interaction
is specified by the  factor $\sum_n t^\ast_{\ka\alpha n} t_{\ka'\alpha'n}$
for which we can obtain analytic expressions from tight-binding theory. 
For $\ka, \ka' \to 0$ and $\alpha = \alpha'=p_x, p_y, p_z$ we obtain
\be
\sum_n t^\ast_{0\alpha n} t_{0\alpha n}
  = \frac{16}{27}\,[3\pds^2\!-4\sqrt{3}\,\pds\pdp
  + 4\pdp^2] .
\ee
Restoring the prefactor from Eq.~(\ref{2.Htrans2}) we find a microscopic
expression for the envelope-function exchange constant $J_{\mathrm{pd}}$.
By including the \emph{full} $(\ka,\ka')$ dependence we recover spatial
anisotropies neglected in that theory. 

Since both denominators in $1/\Delta$ must be negative for $(5,5/2)$ to
be the isolated-ion ground
state, the exchange interaction is \emph{antiferromagnetic},
$J_{\mathrm{pd}}<0$. $|J_{\mathrm{pd}}|$ is
minimized and the effective model has  the widest range of validity when
the $d^5\!\to\!d^4$ and $d^5\!\to\!d^6$ transition energies bracket the
Fer\-mi energy $E_F$ symmetrically. In this case
$J_{\mathrm{pd}} = -48\:\mathrm{meV\,nm}^3$, close to the experimental
value in (Ga,Mn)As \cite{Ohno}. We consider this case in what follows. The
expression for $J_{\mathrm{pd}}$, combined with materials trends
\cite{ZungerSSP}, suggests that $T_c$ of  ${\rm Ga}_{1-x}{\rm Mn}_{x}{\rm
As}_{1-y}{\rm P}_{y}$ quaternary alloys might \emph{increase} with $y$ since
their $d^{5}\!\to\!d^{4}$ transition energy will approach $E_F$, increasing
the value of $J_{\mathrm{pd}}$.  
  
We employ the full $(\ka,\ka')$-dependent hole-im\-pu\-ri\-ty exchange to
evaluate the RKKY interaction between two Mn spins at $0$ and $\mathbf{R}$
and perform the CPT as above. Integrating out the band electrons and
expanding the action to second order in impurity spins we obtain
\ba
\lefteqn{ H_{\mathrm{RKKY}} \: = \: \frac{1}{4\beta\Delta^2}\,
  \sum_{\mu\nu} S_1^\mu S_2^\nu\, \frac{1}{{\cal N}^2}
  \sum_{\ka,\ka'} \sum_{i\omega} \mathrm{Tr}\:
  e^{i(\ka-\ka')\cdot\mathbf{R}} } \nonumber \\
& & {}\times (-i\omega + \hat\epsilon(\ka) - \mu)^{-1}\,
  \hat j^\mu(\ka,\ka')\, (-i\omega + \hat\epsilon(\ka') - \mu)^{-1}
  \nonumber \\
& & {}\times \hat j^\nu(\ka',\ka) \;\, \equiv \;\, -\sum_{\mu\nu}
  J_{\mu\nu}(\mathbf{R})\, S_1^\mu S_2^\nu ,
\label{4.HRKKY1}
\ea
where $\hat\epsilon(\ka)$ is the tight-binding Hamiltonian with ma\-trix
elements $\epsilon_{\alpha'\sigma';\alpha\sigma}(\ka)$ and
$j^\mu(\ka,\ka')_{\alpha'\sigma';\alpha\sigma} \equiv \sum_n
t^\ast_{\ka\alpha n} t_{\ka'\alpha' n}\, \sigma^\mu_{\sigma'\sigma}$. The
trace in Eq.~(\ref{4.HRKKY1}) is over orbital and spin indices. We
diagonalize $\hat\epsilon(\ka) = \hat U^\dagger_{\ka}\, \hat d(\ka)\, \hat
U_{\ka}$, where $\hat d(\ka)$ is the diagonal matrix of band energies
$d_{\alpha\sigma}(\ka)$, and perform the Matsubara sum. It is useful to
express $J_{\mu\nu}(\mathbf{R}) = \int {d^3q}/{(2\pi)^3}\:
e^{i\q\cdot\mathbf{R}}\, J_{\mu\nu}(\q)$
in terms of its Fourier transform.
Making use of the symmetries of $\hat d$ and $\hat U$ we obtain
\ba
\lefteqn{ J_{\mu\nu}(\q) = \frac{v_{\mathrm{uc}}^2}{2\Delta^2}
  \sumk \sum_{\alpha\sigma} f_{\ka\alpha\sigma}
  \sum_{\:\alpha'\sigma'} (1-f_{\ka-\q,\alpha'\sigma'}) } \nonumber \\
& & {}\times \frac{1}{d_{\alpha'\sigma'}(\ka-\q) -
  d_{\alpha\sigma}(\ka)} \:
  [\hat U_{\ka} \hat j^\mu(\ka,\ka-\q)
    \hat U^\dagger_{\ka-\q}]^{}_{\alpha\sigma;\alpha'\sigma'}
  \nonumber \\
& & {}\times [\hat U_{\ka-\q} \hat j^\nu(\ka-\q,\ka)
    \hat U^\dagger_{\ka}]^{}_{\alpha'\sigma';\alpha\sigma} ,
\label{4.Jmnq5}
\ea
where $v_{\mathrm{uc}}$ is the unit-cell volume
and $f_{\ka\alpha\sigma}$ is a Fermi
factor. In the following, we take the electrons to be at $T=0$. We remark
that Eq.~(\ref{4.Jmnq5}) is unreliable when 
$\q$ is comparable to Brillouin-zone dimensions because the band
eigenenergies are then as far from the Fermi energy as
the d-quasiparticle levels. Correspondingly the results for
$J_{\mu\nu}(\mathbf{R})$ are quantitatively reliable only
for $R\gg a$, where $a$ is the dimension of the fcc unit cell.

\begin{figure}[h]
\includegraphics[width=3.25in,clip]{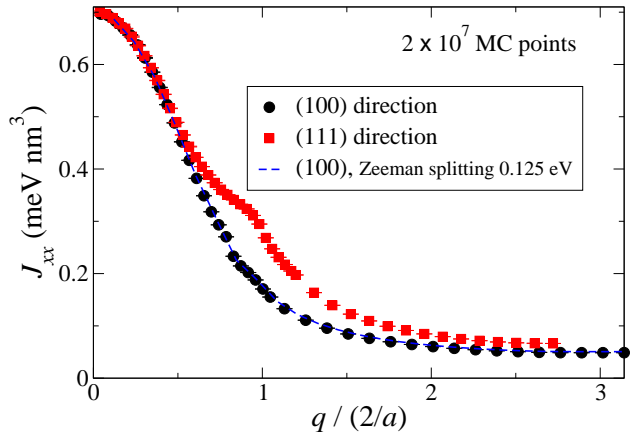}
\caption{(color online) Fourier-transformed RKKY interaction
$J_{xx}(\q)$ and numerical errors
in the (100) and (111) directions for
$E_F=-0.307\:\mathrm{eV}$ relative to the valence-band top, correponding to
a hole concentration of $3.8\times 10^{20}\:\mathrm{cm}^{-3}$.
The dashed curve shows $J_{xx}(\q)$ in the
(100) direction calculated with a band Zeeman splitting of
$0.125\:\mathrm{eV}$,
corresponding to 5\% Mn substitution and full polarization of Mn moments.}
\label{fig.1}
\end{figure}

We have evaluated $J_{\mu\nu}(\q)$ using Monte Carlo (MC) integration with
the {\scshape Vegas} algorithm \cite{VEGAS}. Figure \ref{fig.1} shows
$J_{xx}(\q)$ in the (100) and (111) directions.
At a nonzero Mn density, the interactions between spins are dominated by
the pairwise RKKY interaction only if the mean hole-impurity
exchange interaction is weak \cite{DOM00,KSJ03}. This is indeed the case since
Fig.~\ref{fig.1} shows that the effect of a realistic Zeeman splitting on
$J(\q)$ is small. We note that $J_{\mu\nu}(\q=0)$ is
isotropic; this limit determines the
bulk magnetic anisotropy \cite{AJB01,DOM01}
which vanishes in the present approximation \cite{magneticaniso}.

\begin{figure}[h]
\includegraphics[width=3.25in,clip]{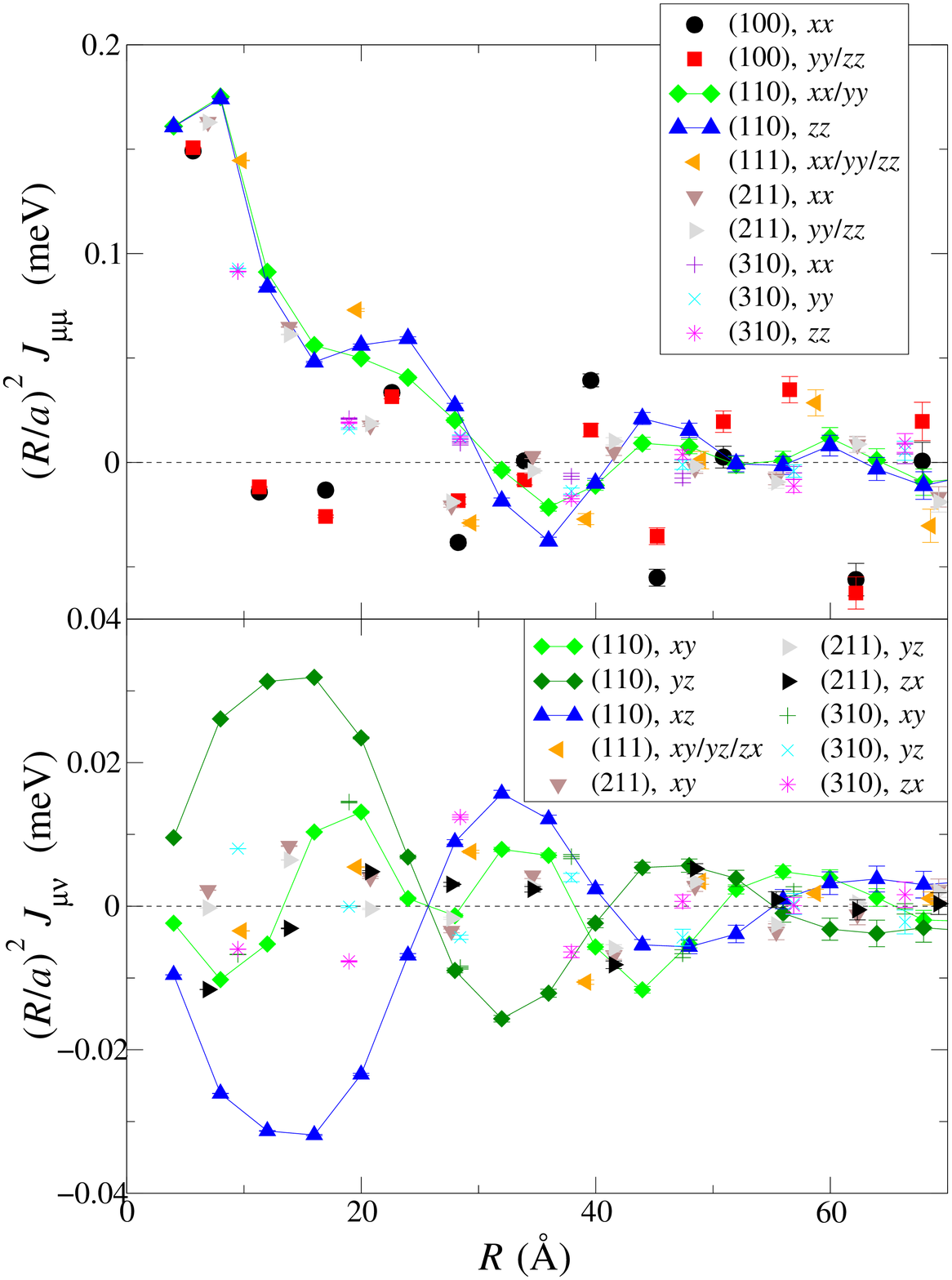}
\caption{(color online)
(a) Diagonal and (b) off-diagonal components of the RKKY
interaction $J_{\mu\nu}(\mathbf{R})$ in various
crystal directions, scaled by $(R/a)^2$.
All results have been obtained with $n_k=36$
and $2\times 10^5$ MC points for each $\q$ point except for
$(qa/2)^2\le0.5$, when $2\times 10^6$ points have been used.
The off-diagonal components vanish exactly along (100).}
\label{fig.2}
\end{figure}

$J_{\mu\nu}(\mathbf{R})$ is evaluated as a Fourier sum over
$J_{\mu\nu}(\q)$ calculated on a cubic grid with $(2n_k)^3/2$ points in the
fcc Brillouin zone, making use of all symmetries. The resulting RKKY
interaction is plotted in Fig.~\ref{fig.2}. It is ferromagnetic at small
separations, as expected. The near-neighbor interactions are not reliable,
both because their  evaluation stretches the validity of the CPT and
because we neglect the \emph{superexchange} interaction, which appears at
fourth order in $H_{\mathrm{hyb}}$, and in which an electron hops virtually
from a Mn d-orbital to a d-orbital on a neighboring Mn site via an
intervening As p-orbital. For larger separations $J_{\mu\nu}(\mathbf{R})$
shows typical Friedel oscillations.

We find a very strong anisotropy in \emph{real} space;
$J_{\mu\nu}(\mathbf{R})$ depends on the direction of $\mathbf{R}$ for
similar $R=|\mathbf{R}|$. This is a consequence of both the directionality 
associated with $pd$ hybridization and of the anisotropy of the band
structure and the Fermi surface; neither effect is included in the
spherical model of Ref.~\cite{ZaJ02}. In Ref.~\cite{BrG03} the real-space
anisotropy is concluded to be small, based on the interaction between two
spins at neighboring sites. For small $R$ we also find relatively weak
anisotropies but at larger $R$ this conclusion does not hold. The isotropic
Gaussian ansatz for the hole-impurity exchange interaction \cite{BrG03}
contributes to this small anisotropy.

The anisotropy in \emph{spin} space, \textit{i.e.}, the deviation of
$J_{\mu\nu}(\mathbf{R})$ from $J(\mathbf{R})\,\delta_{\mu\nu}$, is also
large. For small spin-orbit coupling, the differences between diagonal
components are of second order in spin-orbit coupling, whereas the
off-diagonal components are linear. Only for the smallest separations is 
the relative anisotropy below 10\% as found in Ref.~\cite{BrG03}. At larger
$R$ the anisotropy becomes quite pronounced, as in Ref.~\cite{ZaJ02}.

When the anisotropies are neglected, the moments are fully aligned in the
ground state.
To determine whether or not 
the anisotropies substantially alter the character of the ground state,
we start from a fully aligned (in the \textit{z} direction) spin
configuration and consider the mean effective fields acting on individual
spins,
$H_\mu(\mathbf{R}_i) = S \sum_{j\neq i} J_{\mu z}(\mathbf{R}_i-\mathbf{R}_j)$,
where the sum is over Mn impurity sites. Assumming that the 
Mn ions are distributed completely at random \cite{TSO02,Trev},
the average over all sites
is $\overline{H}_\mu = (xS/v_{\mathrm{uc}})\, J_{\mu z}(\mathbf{q}=0)
\propto \delta_{\mu z}$. On 
average the effective fields align with the average moment, but
spatial fluctuations reduce the overall degree of 
spin polarization. The typical angle of the Mn tilt at a given site
is proportional to the ratio of the \textit{xy} plane effective-field 
components to $\overline{H}_z$. We find
\be
\frac{\overline{H_x^2}}{(\overline{H}_z)^2} 
=  (x^{-1}-1)\, v_{uc} \int \frac{d^3q}{(2\pi)^3}
  \frac{|J_{xz}(\mathbf{q})|^2}{J_{zz}^2(\mathbf{q}=0)} .
\label{allan:2}
\ee
Thus the anisotropies become more important for small Mn fractions $x$.
For the parameters used above we get
$\overline{H_x^2}/[\overline{H}_z]^2 = 3.1\times 10^{-5}\,(x^{-1}-1)$. We
conclude that the anisotropies do not cause a large moment
suppression in (Ga,Mn)As even for $x\sim 0.01$, despite the large
anisotropies. The effect is small because
many moments contribute to the effective
field due to the long-range interaction, averaging over
the anisotropies.
We neglect the indirect
influence of charge scattering, as well as Coulomb interactions and local
chemical shifts. These will reduce the RKKY interaction
at large separations and further reduce the importance of
frustrating interactions \cite{PHS03}. 

To conclude, we have used a Slater-Koster tight-binding model of III-V DMS
to calculate the full momentum dependence of the hole-im\-pu\-ri\-ty
exchange interaction. We find that this interaction depends crucially on
the position of the Mn d-levels relative to the valence band and suggest
that quaternary ${\rm Ga}_{1-x}{\rm Mn}_{x}{\rm As}_{1-y}{\rm P}_{y}$
alloys  might have higher transition temperatures than  ${\rm Ga}_{1-x}{\rm
Mn}_{x}{\rm As}$. Starting from the hole-impurity interaction, we have
calculated the hole-me\-dia\-ted RKKY interaction between impurity spins.
This interaction is highly anisotropic in real and spin space. The
anisotropy crucially depends on two factors partly ignored in previous
works: the nonlocal form of the hole-impurity exchange interaction and the
highly anisotropic band structure. However, despite the strong anisotropies
the local-moment suppression is weak due to the averaging brought about by
the long-range RKKY interaction.

We gratefully aknowledge stimulating discussions with W. A. Atkinson, L.
Brey, T. Dietl, G. Fiete, T. Jungwirth, P. Kacman, T. Schulthess, J.
Sinova, G. Zar\'and, and A. Zunger. AHM was supported by the Welch
Foundation, by the National Science Foundation under grant DMR-0115947 and
by the DARPA SpinS program.

\end{document}